\pdfoutput=1
\documentclass[a4paper,british,floatfix,pdftex,superscriptaddress,%
aps,prl,
reprint,twocolumn,
]{revtex4-2}
\usepackage{amsfonts,amsmath,amssymb}
\usepackage{graphicx}
\usepackage[T1]{fontenc}
\usepackage[utf8]{inputenc}
\usepackage[unicode]{hyperref}
\usepackage{hypernat}
\usepackage[ams,todos]{CMImacros}

\graphicspath{{./}} 
\newcommand*{\methods}{\hyperref[sec:methods]{Methods}\xspace}%

\newcommand{\cfeldesy}{\affiliation{Center for Free-Electron Laser Science CFEL,
    Deutsches Elektronen-Synchrotron DESY, Notkestrasse 85, 22607 Hamburg, Germany}}%
\newcommand{\uhhcui}{\affiliation{Center for Ultrafast Imaging, Universität Hamburg, Luruper
    Chaussee 149, 22761 Hamburg, Germany}}%
\newcommand{\uhhphys}{\affiliation{Department of Physics, Universität Hamburg, Luruper Chaussee 149,
    22761 Hamburg, Germany}}%
\newcommand{\uhhchem}{\affiliation{Department of Chemistry, Universität Hamburg,
    Martin-Luther-King-Platz 6, 20146 Hamburg, Germany}}%
\newcommand{\auchem}{\affiliation{Department of Chemistry, Aarhus University, Langelandsgade 140,
    8000 Aarhus C, Denmark }}%
\newcommand{\saclay}{\affiliation{LIDYL, CNRS, CEA, Université Paris–Saclay, 91191 Gif-sur-Yvette, France}}

\newcommand{\stemail}{\email[Email:~]{sebastian.trippel@cfel.de}}%
\newcommand{\cmiweb}{\homepage[\\Website:~]{https://www.controlled-molecule-imaging.org}}%

\newcommand*{\pyrrolei}{\ensuremath{\pyrrole^+}\xspace}
\newcommand*{\pyrroledii}{\ensuremath{\pyrrole^{2+}}\xspace}
\newcommand*{\pyri}{\ensuremath{\pyr^+}\xspace}
\newcommand*{\pyrdii}{\ensuremath{\pyr^{2+}}\xspace}

\newcommand*{\pyrrolewi}{\ensuremath{\pyrrole(\HHO)^+}\xspace}
\newcommand*{\pyrrolewdii}{\ensuremath{\pyrrole(\HHO)^{2+}}\xspace}

\newcommand*{\pyrwi}{\ensuremath{\pyr(\HHO)^+}\xspace}

\newcommand*{\HHOi}{\ensuremath{\HHO^+}\xspace}

\newcommand*{\hydron}{\ensuremath{\text{H}_3\text{O}^+}\xspace}
\newcommand*{\pyrmhi}{\ensuremath{\text{C}_4\text{H}_4\text{N}^+}\xspace}

\newcommand*{\Ndet}{\ensuremath{N_\mathrm{det}}\xspace}
\newcommand*{\Nshots}{\ensuremath{N_\mathrm{shots}}\xspace}
\newcommand*{\Rdet}{\ensuremath{R_\mathrm{det}}\xspace}

\newcommand*{\Nreal}{\ensuremath{N_\mathrm{real}}\xspace}

\begin{document}
\title{Water is a radiation protection agent for ionised pyrrole}%
\author{Melby Johny}\cfeldesy\uhhcui\uhhphys%
\author{Constant A.\ Schouder}\auchem\saclay
\author{Ahmed Al-Refaie}\cfeldesy%
\author{Lanhai He}\cfeldesy%
\author{Joss~Wiese}\cfeldesy\uhhcui\uhhchem
\author{Henrik Stapelfeldt}\auchem%
\author{Sebastian Trippel}\stemail\cmiweb\cfeldesy\uhhcui%
\author{Jochen Küpper}\cfeldesy\uhhcui\uhhphys\uhhchem%
\date{\today}%
\begin{abstract}\noindent%
   Radiation-induced damage of biological matter is an ubiquitous problem in nature. The influence
   of the hydration environment is widely discussed, but its exact role remains elusive. Utilising
   well defined solvated-molecule aggregates, we experimentally observed a hydrogen-bonded water
   molecule acting as a radiation protection agent for ionised pyrrole, a prototypical aromatic
   biomolecule. Pure samples of pyrrole and \pyrrolew were outer-valence ionised and the subsequent
   damage and relaxation processes were studied. Bare pyrrole ions fragmented through the breaking
   of C--C or N--C covalent bonds. However, for \pyrrolewi, we observed a strong protection of the
   pyrrole ring through the dissociative release of neutral water or by transferring an electron or
   proton across the hydrogen bond. Overall, a single water molecule strongly reduces the
   fragmentation probability and thus the persistent radiation damage of singly-ionised pyrrole.
\end{abstract}
\maketitle%

\section{Introduction}
\label{sec:introduction}
The damage of biological matter upon the interaction with UV
radiation~\cite{Crespo-Hernandez:CR104:1977} or ionising radiation~\cite{Lehnert:BioIonRad:2007},
such as x-rays~\cite{Lehnert:BioIonRad:2007, Alizadeh:ARPC66:379},
$\gamma$-rays~\cite{Ito:IJRB63:289}, and $\alpha$-particles~\cite{Kim:PNAS108:11821}, or other
charged particles~\cite{Alizadeh:ARPC66:379, Boudaiffa:Science287:1658, Mark:NatPhys6:82} is a major
environmental impact on living organisms~\cite{Lehnert:BioIonRad:2007, Crespo-Hernandez:CR104:1977}.
For instance, inner-shell-, inner-valence-, or outer-valence-ionised states can relax in various
pathways that form cationic species, which can result in break up of the
molecules~\cite{Jahnke:JPB48:082001, Cederbaum:PRL79:4778,Alizadeh:ARPC66:379}. One highly relevant
mechanism of DNA-strand breaks is \emph{via} autoionisation or excitation caused by low-energy
secondary electrons~\cite{Simons:ACR39:772, Alizadeh:ARPC66:379, Boudaiffa:Science287:1658,
  Martin:PRL93:068101, Mark:NatPhys6:82}.

Regarding the radiation damage of molecules, ionisation and excitation are similar: In both cases,
vacancies are created in the occupied molecular orbitals, and this can lead to bond breaking. In the
case of ionisation, the electron is directly transferred to the continuum, leaving the molecular ion
behind, while excitation may result in the population of dissociative excited
states~\cite{Ng:VUV:1991}. Typical sources for single ionisation of biological matter in aqueous
environments are deep UV radiation or the interaction with radicals, slow electrons, or
ions~\cite{Alizadeh:ARPC66:379, Ng:VUV:1991, Garrett:CR105:355, Lehnert:BioIonRad:2007}. While deep
UV radiation is efficiently blocked by the earth's atmosphere~\cite{Lubin:Nature377:710} it is
omnipresent in outer space~\cite{Barstow:ExtremeUVAstro}. Harder, \eg ionising, radiation
penetrates the atmosphere.

Molecular assemblies such as clusters, droplets, and even large molecules like proteins in their
natural solvation environment are known to allow for additional relaxation pathways due to
intermolecular interactions~\cite{Schultz:Science306:1765, Fang:nature462:200, Marx:Nature397:601,
   Golan:NatChem4:323, Horke:PRL117:163002, Mudrich:NatComm11:112, Slavicek:JPCL7:234,
   Jahnke:JPB48:082001, Barbatti:PNAS107:21453, Zobeley:JCP115:5076}. These pathways may lead to a
protection of the molecule especially if the biomolecule is directly affected by the
radiation~\cite{Crespo-Hernandez:CR104:1977}. On the other hand, secondary species originating from
the ionisation of surrounding solvent molecules can induce new pathways that lead to biomolecular
destruction.

Hydrogen-bonded solute-solvent complexes allow for quantitative investigations of these
effects~\cite{Ren:NatPhys:79:1745, Jahnke:NatPhys6:139, Xu:ACIE57:17023, Richter:NatComm9:4988}. One
of the important electronic relaxation channels of such solvated-molecule clusters after x-ray
ionisation, electron-impact ionisation, or $\alpha$-particle irradiation was ascribed to
intermolecular Coulombic decay (ICD)~\cite{Jahnke:JPB48:082001, Cederbaum:PRL79:4778,
   Ren:NatPhys:79:1745, Xu:ACIE57:17023, Jahnke:NatPhys6:139, Mucke:NatPhys6:143}. This resulted in
the formation of mainly charge-separated di-cationic complexes which undergo fragmentation
\emph{via} electrostatic repulsion. A competing ultrafast relaxation channel of hydrogen-bonded
complexes after inner-shell ionisation, which may protect biomolecules, is intermolecular electron-
or proton-transfer-mediated charge separation~\cite{Zobeley:JCP115:5076, Slavicek:JPCL7:234,
   Xu:ACIE57:17023}. This was also observed following x-ray ionisation of the water
dimer~\cite{Richter:NatComm9:4988} and liquid water~\cite{Thuermer:NatChem5:590,
   Petr:JACS136:18170}.

To examine the influence of nano-hydration on the dynamics of biomolecules various studies using
mass spectrometric techniques were performed~\cite{Berthias:CPC16:3151, Liu:PRL97:133401,
   Kim:IJMS219:11, Barc:IJMS365:194, Pandey:EPJD71:190, Kim:JPC100:7933, Kocisek:JPCL7:3401,
   Wang:NatComm11:2194, Sukhodub:CR87:589, Neustetter:JASMS28:866}. These experiments can
essentially be divided into three approaches: In the first approach one uses mass selective ion beam
methods to extract individual clusters in the case of cationic and anionic
samples~\cite{Berthias:CPC16:3151, Liu:PRL97:133401, Comte:JPCA3:775}. The initialisation of the
dynamics, \eg ionisation or collisions, is species unspecific which results in an undefined initial
condition. For instance, it is not possible to avoid ionisation of the water as the initial step.
The second approach of experiments utilises a localised initiation for the dynamics inside the
cluster using, \eg multi-photon ionisation~\cite{Kim:IJMS219:11, Barc:IJMS365:194,
   Pandey:EPJD71:190} or slow electrons~\cite{Kim:JPC100:7933, Kocisek:JPCL7:3401,
   Wang:NatComm11:2194}. Here, however, experiments are done with mixed samples, \eg various
cluster sizes and isomers present in the molecular beam simultaneously. An unambiguous assignment of
the fragments to a corresponding parent cluster is not possible. This is especially a problem in the
case where the isolated molecules and the clusters give rise to the same fragments. The third
approach utilises neither a well-defined probe nor a localised trigger~\cite{Sukhodub:CR87:589,
   Neustetter:JASMS28:866}.

A key ingredient for the indirect destruction pathways of biological matter is the radiolysis of
water~\cite{Loh:Science367:179, Svoboda:SciAdv6:eaaz0385, Garrett:CR105:355}, likely because
typically $\ordsim3/4$ of a cell's volume comprises an aqueous
environment~\cite{Alizadeh:ARPC66:379, Garrett:CR105:355}. In this case, reactive cations, radicals,
anions, or electrons are produced inside the water environment, which can trigger biomolecular
fragmentation~\cite{Lehnert:BioIonRad:2007, Michael:Science187:1603, Loh:Science367:179,
   Svoboda:SciAdv6:eaaz0385, Wang:JACS131:11320, Bao:PNAS103:5658}. In this context, numerous
experimental studies claim that the hydration environment can either inhibit or enhance
radiation-induced biological damage~\cite{Ma:SciAdv3:1, Ma:NatComm10:102, Gu:CR112:5603,
   Bao:PNAS103:5658, Smyth:JCP140:184313, Ito:IJRB63:289, Kocisek:JPCL7:3401, Garrett:CR105:355,
   Barc:IJMS365:194, Liu:PRL97:133401, Milosavljevic:jpcl5:1994}. This includes a recent study
showing a relatively weak catalysis effect, \ie an enhancement of radiation damage, due to water
being present~\cite{Wang:NatComm11:2194}. The details obviously depend on the nature of the ionising
radiation and the specific potential energy landscape of the biomolecule in its hydration
environment~\cite{Lehnert:BioIonRad:2007, Alizadeh:ARPC66:379}.

Pyrrole, a heterocyclic aromatic molecule, is a UV-absorbing chromophore, \eg in hemes and
chlorophylls~\cite{Ulijasz:Nature463:250}. Pyrrole is also a subunit of indole, 3-methylindole, and
tryptophan, which are of great relevance as the principal UV absorbers of
proteins~\cite{Ashfold:Science1637:1640, Sobolewski:PCCP4:1093}. The photophysical and photochemical
properties of indole and pyrrole are sensitive to the hydration
environment~\cite{Sobolewski:CP259:181, Lippert:CPL376:40, Sobolewski:CPL321:479,
   Korter:JPCA102:7211, Pal:PNAS99:1763, Onvlee:NatComm13:7462}: upon UV absorption these
chromophores indirectly populate an excited $^1\pi\sigma^*$ state, which is repulsive along the
N-H-stretching coordinate~\cite{Ashfold:Science1637:1640, Sobolewski:PCCP4:1093,
   Sobolewski:CP259:181}. This triggers an ultrafast internal-conversion process to the ground
state, essential for the photostability of proteins~\cite{Domke:NatChem5:1755}.

The \pyrrolew cluster has a well-defined structure with a hydrogen bond between pyrrole's N-H site
and the water's oxygen~\cite{Tubergen:JPC97:7451}. This can, due to its similarities with
indole-water, reflect the strongest interaction between pyrrole and surrounding \HHO in aqueous
solution~\citep{He:JPCL14:10499}. H-elimination dynamics from the N-H site of pyrrole, mediated by
the electronic excitation of the $^1\pi\sigma^*$ state~\cite{Ashfold:Science1637:1640,
  Roberts:FD95:115, Kirkby:CPL683:179, Lippert:CPC5:1423} or by vibrationally mediated
photodissociation~\cite{Grygoryeva:aipadv9:0351511}, was studied by time-resolved photoion and
photoelectron spectroscopy. Theoretical calculations for electronically excited \pyrrolew clusters
predicted electron transfer across the hydrogen bond without photodissociation of the pyrrole
moiety~\cite{Frank:CP343:347, Sobolewski:CPL321:479}. Direct ionisation from the HOMO and HOMO-1
orbitals of pyrrole~\cite{VanDenBrom:PCCP7:892,Rennie:CP250:217} and the HOMO orbital of
\pyrrolew~\cite{Schuetz:PCCP19:3970, Johny:CPL721:149} does not lead to fragmentation of the
aromatic ring. However, the dissociation from excited cationic states of pyrrole led to molecular
fragmentation~\cite{VanDenBrom:PCCP7:892,Rennie:CP250:217}, and similar processes are expected for
\pyrrolew.

Here, we experimentally investigated the damage incurred in singly- and doubly-ionised pyrrole
molecules and the effect of solvation by comparing the fragmentation pathways of bare pyrrole and
microsolvated \pyrrolew heterodimers using the combination of pure samples of either
species~\cite{Johny:CPL721:149} and local, site-specific multi-photon ionisation. Our clean approach
enables the systematic investigation of the role of water solvent on the photophysics of pyrrole and
for hydrated biomolecules in general.

\begin{figure}
   \includegraphics[width=\linewidth]{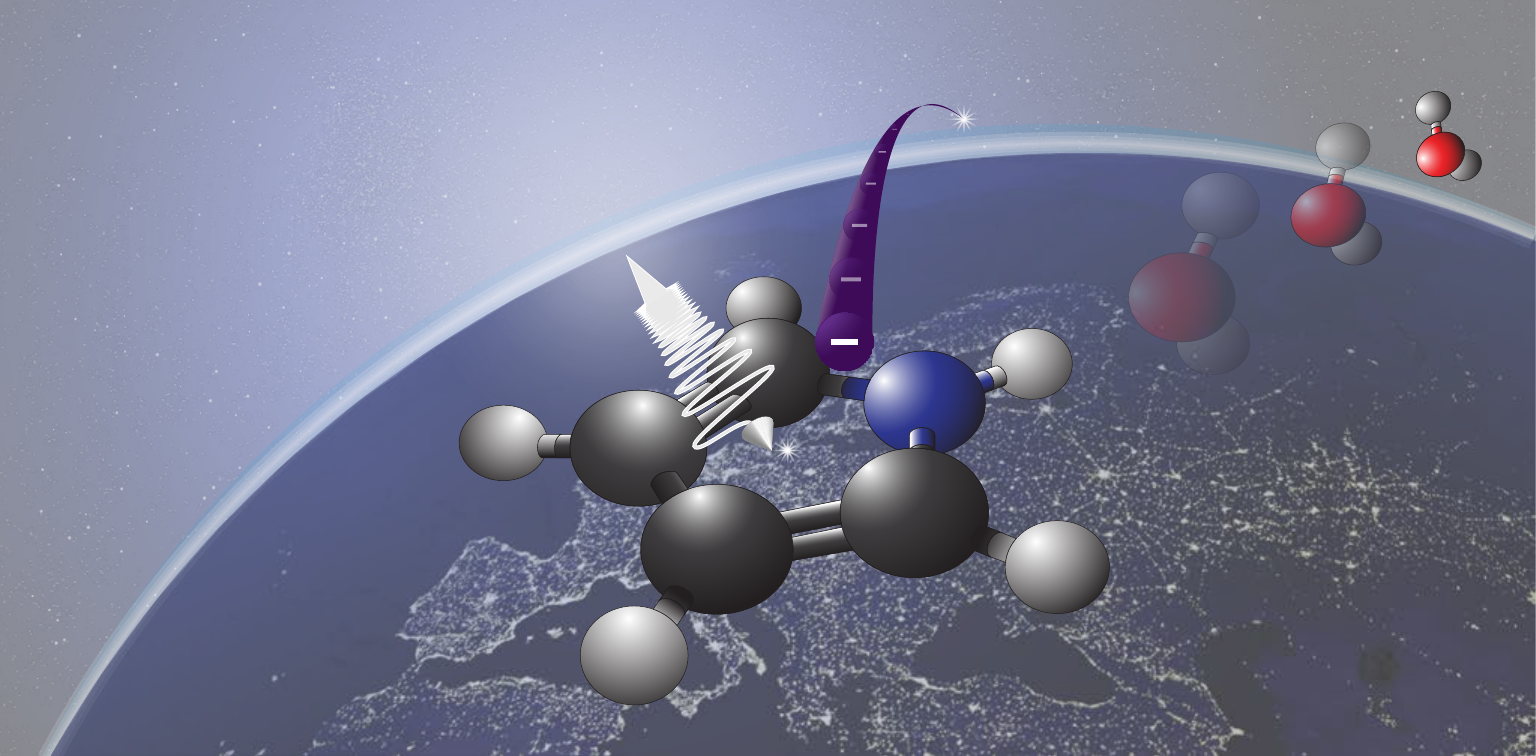}%
   \caption{Schematic representation of the ionisation scheme and radiation-protection mechanism in
      \pyrrolew. Single ionisation of \pyrrolew is followed by its dissociation, with the leaving
      neutral water molecule allowing the aromatic ring to stay intact without further
      fragmentation.}
   \label{fig:TOC_fig}
\end{figure}

A schematic representation of our ionisation strategy is represented in \autoref{fig:TOC_fig}.
We mimicked the triggering of radiation damage through outer valence ionisation as it would
naturally occur in secondary processes such as the interaction with slow electrons or UV photons.
Pyrrole and \pyrrolew were site-specifically ionised by removing electrons from the highest-energy
molecular orbitals, whose densities are predominantly localised on the aromatic ring as shown in
\autoref{fig:orbital}. The site-specific ionisation allows a direct comparison of the subsequent
fragmentation of the two systems since the ionisation process is not significantly altered.
Ionisation was achieved by strong-field ionisation (SFI) using 800~nm laser pulses with a peak
intensity of $10^{14}~\Wpcmcm$ and a pulse duration of 30~fs, see \methods for details. On one hand,
valence ionisation of bare pyrrole resulted in extensive fragmentation, \ie breaking of the aromatic
ring. On the other hand, for singly-ionised \pyrrolew we mainly observed solely the breaking of the
hydrogen bond with the water molecule leaving the intact pyrrole ring. In this case, breaking of the
actual biomolecule is strongly suppressed.

\section{Results}
\label{sec:results}
Our comparison of the fragmentation dynamics of bare and singly-microsolvated pyrrole built on the
production of very pure molecular beams of pyrrole and \pyrrolew, respectively. The electrostatic
deflector was used to spatially separate the different species by their dipole-moment-to-mass ratio
from a cold molecular beam~\cite{Chang:IRPC34:557, Johny:CPL721:149}; see \methods for details. The
species-selected molecular beam had, in the case of pyrrole, a purity of $\ordsim95$~\%, with a
$\ordsim5$~\% contamination by pyrrole homodimer. The purity of the \pyrrolew beam was
$\ordsim99$~\% with the main contamination being the water dimer~\cite{Johny:CPL721:149}.

\subsection{Fragmentation dynamics of \pyrrolei}
\begin{figure}
   \includegraphics[width=\linewidth]{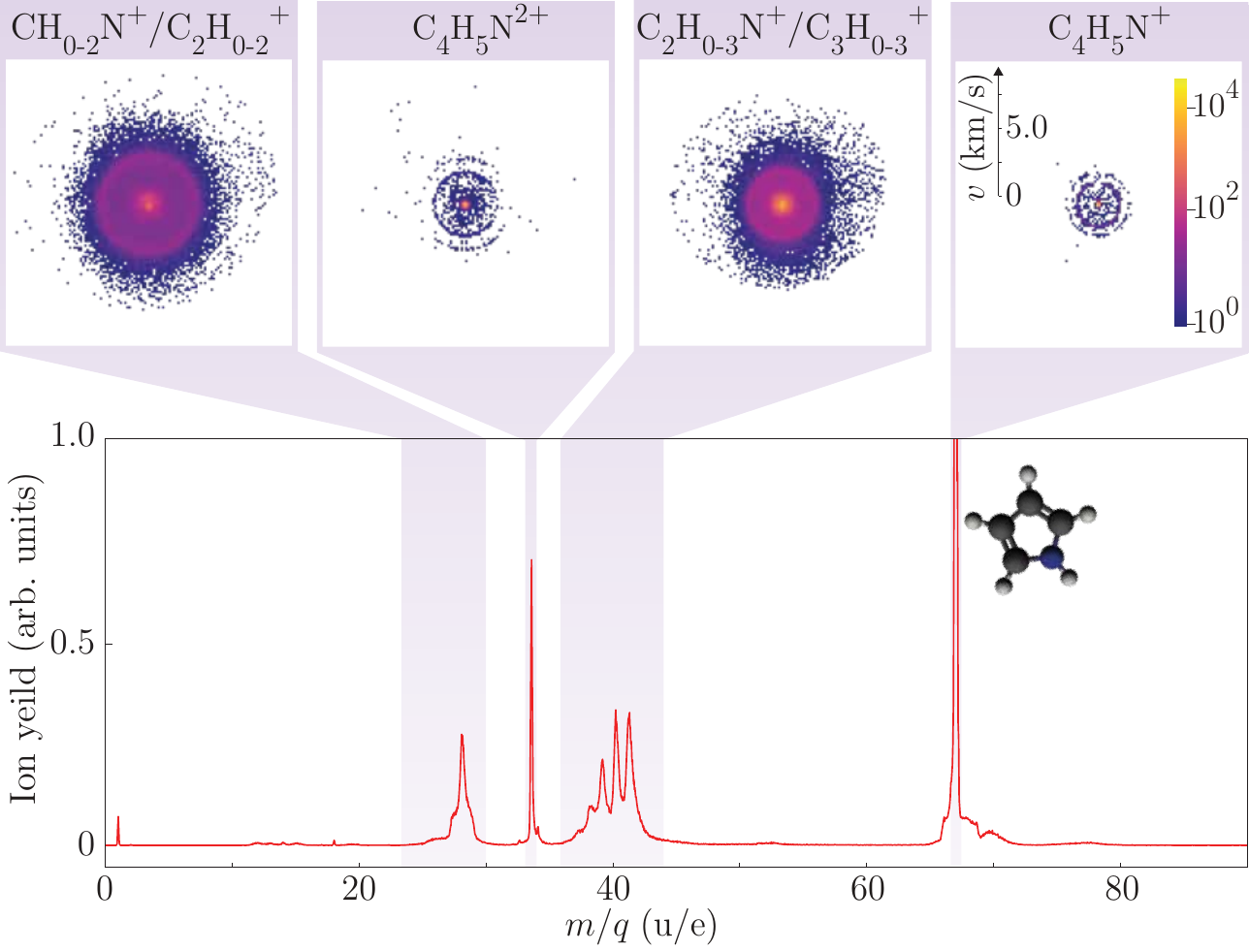}%
   \caption{Time-of-flight mass spectrum and corresponding velocity-map images of the ions generated
      by strong-field ionisation of pyrrole. The structure of pyrrole is given on the right of the
      lower panel. The colourmap and velocity scale holds for all velocity-map images.}
   \label{fig:pyr:tof}
\end{figure}
\autoref{fig:pyr:tof} shows the time-of-flight mass spectrum (TOF-MS) and corresponding velocity-map
images (VMIs) of all ions resulting from strong-field ionisation of pyrrole. All data were recorded
simultaneously using a Timepix3 camera~\cite{Zhao:RSI88:113104, AlRefaie:JINST14:P10003,
   Bromberger:JPB55:144001}. Rings in the VMIs occur for two-body Coulomb explosion, \ie a
charge-repulsion-driven fast breakup into two positively charged fragments. These fragmentation
channels obey momentum conservation in the recoil frame. Low-kinetic-energy (KE) features correspond
to intact parent ions or ions created from dissociative single ionisation. The data shows no
signatures of three-body breakup beyond hydrogen atom/proton loss.

The most prominent feature in the TOF-MS is the narrow peak at \mq{67}, assigned to \pyrrolei, on
top of a broader pedestal. The peak corresponds to the sharp central dot in the corresponding VMI.
The pedestal correlates with the rings in the VMI which are assigned to \pyrrolei from Coulomb
explosion of the pyrrole homodimer. The TOF-MS peak at \mq{33.5} and the central dot in the
corresponding VMI are assigned to doubly-ionised pyrrole. Its pedestal in the TOF-MS and the
corresponding rings in the VMI were attributed to \pyrroledii from Coulomb explosion of
multiply-ionised pyrrole homodimer. In both cases 95~\% of the signal strengths were in the central
peaks, \ie originated from the monomer.

The signals in the mass-to-charge regions \mqr{24}{30} and \mqr{35}{44} correspond to fragments from
the breakup of the pyrrole ring. Some possible ionic products are labeled in \autoref{fig:pyr:tof},
in line with the mass peak assignment after electron-impact- and photo-ionisation of
pyrrole~\cite{VanDenBrom:PCCP7:892, Profant:JPCA111:12477}. A clear assignment of the various
individual peaks observed in the two mass regions to specific fragments was not possible due to
overlapping signals and ambiguities in the construction of specific mass-to-charge ratios out of
possible feasible fragments. The situation was further complicated by the fact that some fragments
lost hydrogen atoms. However, proton loss after double ionisation was ruled out due to the lack of
correlations between the detected protons and the fragments observed in the two mass regions. The
small proton peak in the spectrum was attributed to initial charge states $\larger2$, which are not
further discussed in the current work. The structure at \mq{70} is assigned to contributions from
larger clusters. The expected \mq{68} $^{13}$C-pyrrole isotopologues peak is suppressed due to the
strong parent ion signal, see \methods. Intense central peaks in the VMIs correspond to fragments
from dissociative ionisation of singly-charged pyrrole. Rings in the VMIs, contributing
$\ordsim30$~\% of the total ion count, correspond to fragments from Coulomb explosion of
doubly-charged pyrrole. The Coulomb explosion signals of the \mqr{24}{30} and \mqr{35}{44} mass
regions are correlated for the double ionization of bare pyrrole, as confirmed by
kinetic-energy-selected coincidence maps and momentum conservation.

\subsection{Fragmentation dynamics of \pyrrolewi}
\begin{figure}
   \includegraphics[width=\linewidth]{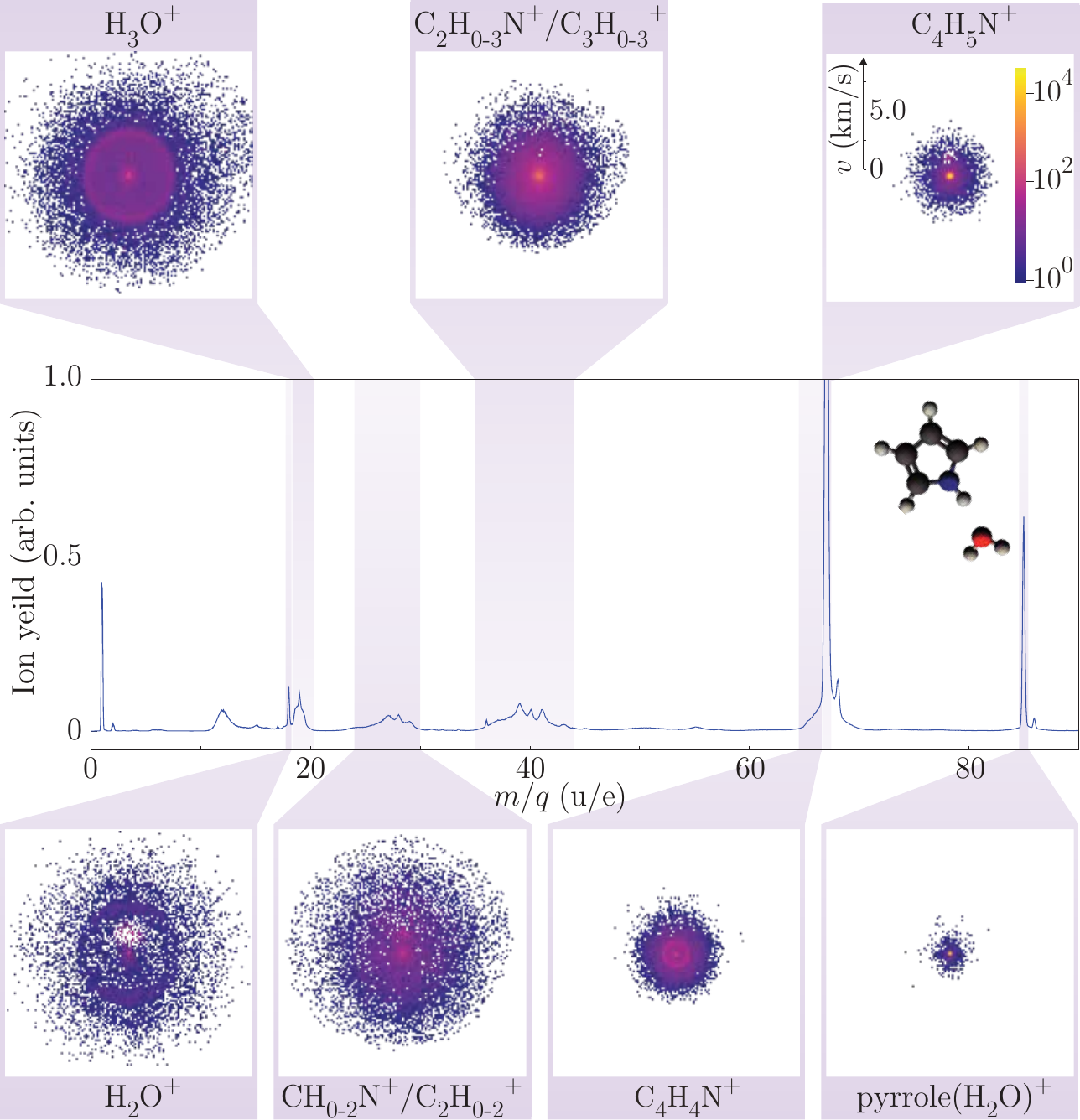}
   \caption{Time-of-flight mass spectrum of \pyrrolew following SFI with the velocity-map images of
      the ions, which were recorded simultaneously after strong-field ionisation. Also shown are the
      structure of the heterodimer, sum formulas of all fragments, and the velocity ranges and
      colourmap for all velocity-map images.}
   \label{fig:pyrwat:tof}
\end{figure}

\autoref{fig:pyrwat:tof} shows the TOF-MS and VMIs of all ions resulting from strong-field
ionisation of \pyrrolew.
Again all data were recorded simultaneously using the Timepix3 detector. All VMIs exhibit a central
low-KE part due to single ionisation as well as sharp or diffuse higher-KE signals.

The peak at \mq{85}, with a central dot in the VMI, corresponds to the \pyrrolew parent ion. The
strongest peak in the TOF-MS is again at \mq{67}, the pyrrole-monomer cation, which resulted from
the dissociation of the hydrogen bond in singly-ionised \pyrrolew. This is confirmed by its broader
KE distribution in the VMI owing to recoil from the momentum conservation with the leaving neutral
water molecule. Furthermore, no correlations between singly charged pyrrole ions themselves were
found in the PIPICO map. We would expect to see this correlation for the higher charge states
present in our data and, therefore, infer negligible contributions from pyrrole homodimers in our
deflected molecular beam in line with previous studies~\cite{Johny:CPL721:149, Bieker:JPCA123:7486}.

The peaks at \mq{66} and \mq{19} correspond to \pyrmhi and \hydron, respectively. Both fragments
exhibit sharp rings in their VMIs, with correlated ions that obeyed momentum conservation,
demonstrating a two-body Coulomb explosion break-up channel of \pyrrolewdii including a proton
transfer from pyrrole to water. A weaker \HHOi channel, $\ordsim3/4$ of which originates from
\pyrrolewdii whereas the remaining signal can be attributed to the water homodimer, shows the direct
charge-separating breakup of \pyrrolewdii across the hydrogen bond.

As for pyrrole, fragments within the regions \mqr{24}{30} and \mqr{35}{44} were detected due to the
breakup of the aromatic ring. However, they showed broad structureless distributions in the VMIs
which were not correlated with each other. The high-KE ions in these regions originate from
\pyrrolewdii after three-body fragmentation processes. These channels involved \hydron as a second
ionic partner, as well as a neutral fragment. The small peak at \mq{36} was attributed to
singly-ionised water homodimer (\HHOd)~\cite{Bieker:JPCA123:7486}.

\section{Discussion}
\label{sec:discussion}
Overall, single- and double-valence ionisation of pyrrole led to a significant breakup of the
aromatic ring, in addition to the formation of singly- or doubly-charged parent ions. Single
ionisation of pyrrole into the $^2\text{A}_{2}$ ground state and the $^2\text{B}_{1}$ first excited
state of the cation led to the formation of a stable parent ion~\cite{VanDenBrom:PCCP7:892,
   Rennie:CP250:217}. Single ionisation also caused fragmentation of the pyrrole moiety through
various dissociative pathways, which resulted in low-kinetic-energy ions. This was due to ionisation
into diffuse bands of excited electronic states of the cation with energies in the range of
12\ldots15~eV, \ie around 4~eV above the ground-state energy of the
cation~\cite{VanDenBrom:PCCP7:892, Rennie:CP250:217, Palmer:CP238:179, Derrick:IJMSIPHYS6:191,
   Willett:JACS102:6774}. With Koopmans' theorem, the energies of these excited states match well
with the ionisation energies of HOMO-2 to HOMO-6 orbitals of the \pyrrole
molecule~\cite{VanDenBrom:PCCP7:892}. Double ionisation caused fragmentation of the aromatic ring
driven by Coulomb explosion.

Surprisingly, the scenario was very different for singly- and doubly-ionised \pyrrolew. New
relaxation pathways emerged due to the hydrogen bond with the water molecule -- despite the
predominantly localised ionisation with the removal of electrons from \pyrrole's $\pi$ orbitals. For
example, in the case of singly-ionised \pyrrolew, we observed a strong dissociation channel for the
hydrogen bond, \ie the loss of neutral water, which protected the pyrrole ring from fragmentation.
Furthermore, after double ionisation additional Coulomb-explosion channels appeared for \pyrrolewdii
compared to \pyrrolei, such as the \hydron channel with its counter ion \pyrmhi. Both fragments
showed sharp rings in the corresponding VMIs, demonstrating that the pyrrole ring was left intact
with only a proton transferred to the water moiety. Additionally, two body Coulomb-explosion
channels of \pyrrolewdii in the regions \mqr{24}{30} and \mqr{35}{44}, \ie clear rings, were absent
and the uncorrelated signal was much weaker than the strong signals observed for the pyrrole
monomer. In total, the fragmentation pathways were significantly influenced, \ie fragmentation of
the pyrrole ring was strongly reduced by the attached single water molecule.

Single ionisation of \pyrrolew into the ground state of the cation created intact parent ions with
the charge localised on the aromatic moiety. The binding energy of the cationic heterodimer was
determined to 670~meV~\cite{Schuetz:PCCP19:3970}. The energy of the first excited state is expected
to be roughly 1~eV above the ground state of the cation, as in the case of the pyrrole
monomer~\cite{VanDenBrom:PCCP7:892, Palmer:CP238:179}. In the TOF-MS of \pyrrolew after
single-ionization, \autoref{fig:pyrwat:tof}, the dominant peak was \mq{67}, so a dissociation
product of \pyrrolewi. The pyrrole cation arose from the dissociation following single ionisation
of \pyrrolew. Assuming a similar ionisation cross-section for pyrrole and \pyrrolew into the first
electronically excited state of the cation led to the conclusion that this state of \pyrrolewi was
dissociative. Otherwise the \pyrrolew spectrum after single-ionisation would be dominated by
\pyrrolewi. Therefore, the aromatic ring was protected after ionisation into the ground and the
first excited state of the \pyrrolew cation. We also observed similar aromatic ring fragmentation
products for \pyrrolewi, as in the case of \pyrrolei, from the above-mentioned diffuse
dissociation band~\cite{VanDenBrom:PCCP7:892, Palmer:CP238:179}. However, the relative intensities
between stable aromatic rings versus fragmentation channels varied significantly for the monomer and
the heterodimer, leading to a higher ring fragmentation probability for the former.

To unravel the influence of the microsolvation on the radiation-induced damage and to quantify the
degree of protection for ionised pyrrole in the presence of a single water molecule, we normalised
the ion yields for the \pyrrole and \pyrrolew species after SFI with respect to each other. We
normalised both species based on their total single-ionisation ion yield by assuming the
single-ionisation cross-section to be the same. This is justified by the similarity of
laser-intensity-dependent single-ionisation ion yield for the two species as described in the
\methods. Furthermore, the single, double, and higher-order ionisation channels were separated to
provide a direct comparison of the fragmentation yields for both species as a function of the
initial charge state. Corrections for saturation effects in detector areas with high count rates
were statistically taken into account, see \methods. In addition, the fragmentation channels were
classified into ring-fragmenting and ring-protecting channels.

\begin{figure}
   \includegraphics[width=\linewidth]{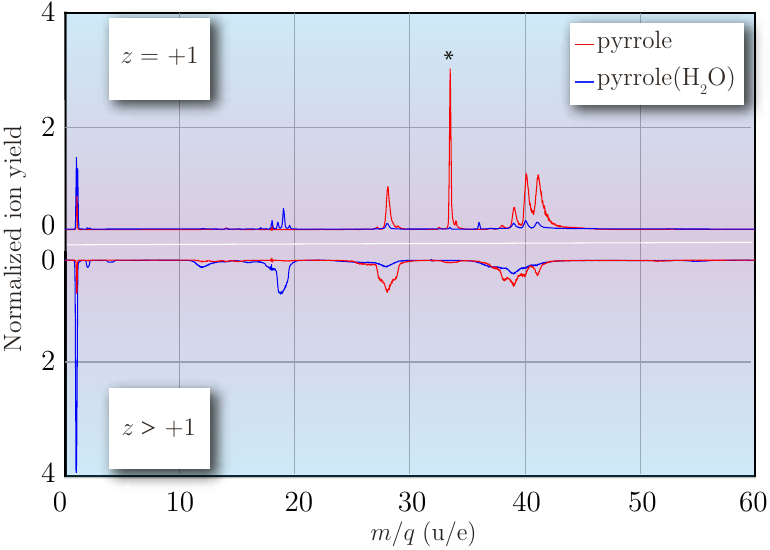}%
   \caption{Comparison of the normalised TOF-MS of pyrrole (red) and \pyrrolew (blue) following SFI.
      The number of charges created on the system after strong-field ionisation is denoted by $z$.
      The upper panel corresponds to an initial charge state $z=+1$. The peak marked with the
      asterisk corresponds to intact \pyrdii from double ionisation of pyrrole. The lower section is
      for the charge states $z>+1$. Signals for \mqr{2}{14} in the lower section originated from
      initial charge states with $z>+2$.}
   \label{fig:compare}
\end{figure}

The momentum maps for the specific mass-to-charge regions were taken into account in order to
separate single, double, and higher-order ionisation channels. Almost all VMIs shown in
\autoref{fig:pyr:tof} and \autoref{fig:pyrwat:tof} had contributions from ions with low as well as
with high kinetic energy. We attribute the low-KE ions to single ionisation and the high-KE ions to
double or higher-order ionisation, respectively. Gating on the momenta with $p<30$~\ukms resulted in
a normalised mass spectrum (NORMS), which is a composition of single-ionisation channels and the
doubly charged parent ion channel that also exhibits low kinetic energy. Gating on the momenta with
$p>30$~\ukms resulted in a NORMS for the higher-order ionisation channels. The resulting normalised
gated time-of-flight mass spectra for pyrrole and \pyrrolew, following SFI, are shown in
\autoref{fig:compare} for \mqr{0}{60}, which contain all ring-breaking fragments.
The upper panel corresponds to the single-ionisation channels ($z=+1$), whereas the lower panel
corresponds to double($z=+2$) or higher($z>+2$) ionisation channels.

Both species showed similar fragmentation products, especially in the mass-to-charge regions
\mqr{24}{30} and \mqr{35}{44}, which were the channels arising from the breaking of the pyrrole
ring. However, they vastly differed in the specific ion yield and we observed a significant
reduction of fragments arising from ring fragmentation for \pyrrolew than for pyrrole after single
ionisation. For both, pyrrole and \pyrrolew, following SFI, the contributions in the NORMS for
$z\ge+2$ in the region \mqr{15}{60} were dominated by double ionisation. This includes the large
differences in the region \mqr{18}{19} where hydronium and water ions are formed. The contributions
from triple ionisation were statistically estimated to less than 5~\% and are thus negligible, see
\methods. Furthermore, the NORMS peaks in the region \mqr{1}{14} originated from charge states with
$z>+2$, confirmed by a covariance analysis. In the case of higher-order ionisation, \ie $z>2$, a
direct comparison of the ion yield of pyrrole and \pyrrolew after SFI from the NORMS was not
feasible due to the complex fragmentation processes and overlapping fragmentation channels.

To estimate the extent of fragmentation protection after single and double ionisation of pyrrole in
a microsolvated environment, the observed fragmentation channels were classified into ring-breakup
and ring-protection channels. Based on this, the ring-fragmentation probability is defined as
$P= N_\mathrm{b}/N_\mathrm{total}$ with the number of fragments where the ring is broken
$N_\mathrm{b}$ and the total ion yield $N_\mathrm{total}$.

First, we considered ring-breakup and ring-protection channels following single ionisation of
pyrrole and \pyrrolew. For pyrrole, the parent ion was considered as a channel leaving the molecule
intact. For single ionisation of \pyrrolew, in addition to the parent ion, the dominant
ring-protection channel was the dissociation of the hydrogen bond, \ie the loss of neutral water.
Furthermore, the dissociative single-ionisation processes resulting in low-KE ions of \HHOi,
\hydron, and \pyrmhi, prevented the aromatic ring from fragmentation. All other low-KE ions in the
mass-to-charge region \mqr{15}{60} were considered as ionic products that originated from the
fragmentation of the aromatic ring. The ring-fragmentation probabilities for pyrrole and \pyrrolew
after single ionisation were then determined by counting the ions in the momentum-map images with a
gating specifically on the low-KE part in the specific mass-to-charge regions. The projection of
ions from higher charge states ($z=+2$) into the center of the momentum-map images was estimated
statistically, see~\methods, and this contribution was subtracted.

\begin{figure}
  \includegraphics[width=0.99\linewidth]{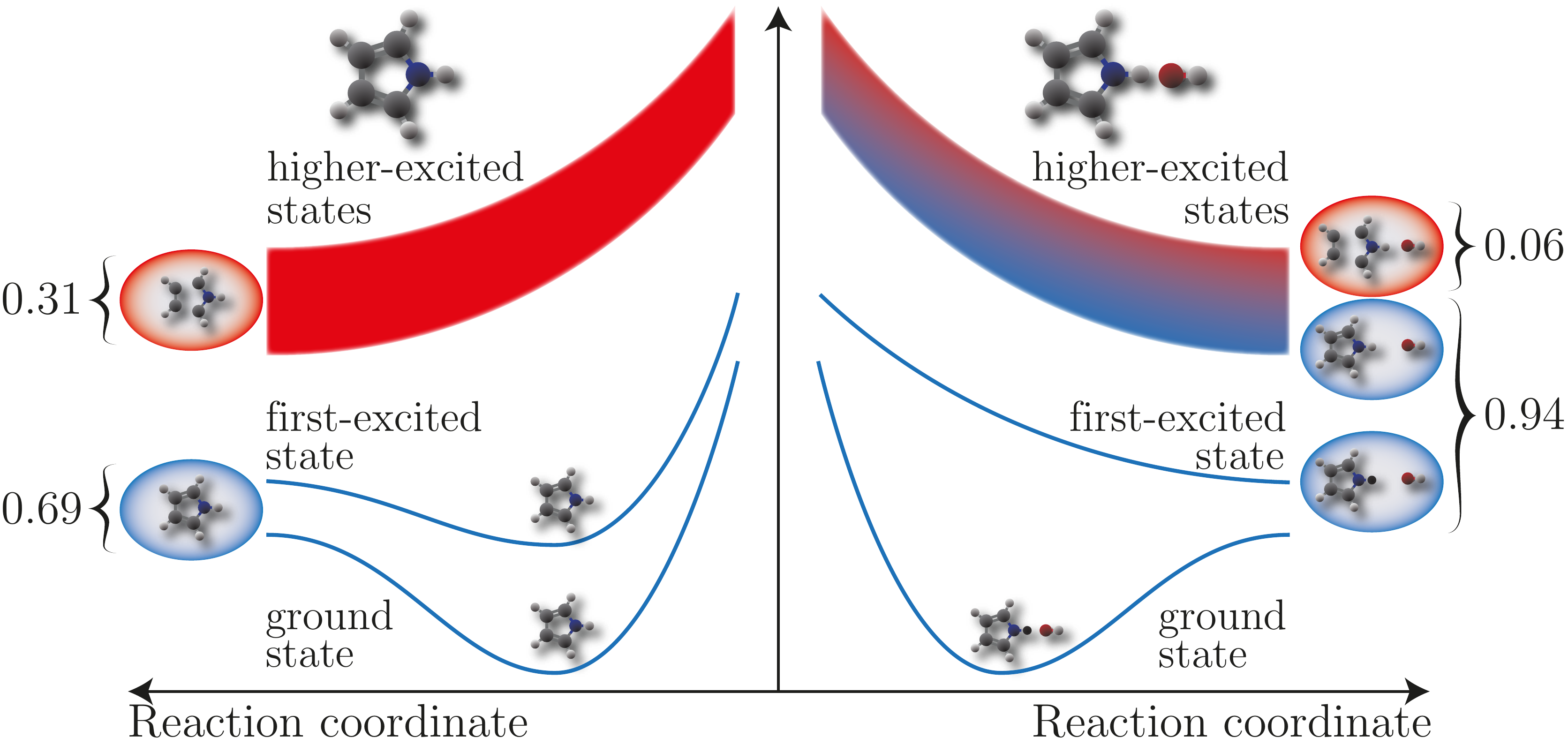}
  \caption{Schematic representation of the potential energy surfaces and the first dissociation band
     of \pyrrolei and \pyrrolewi together with the ring-protection and ring-breaking probabilities.
     Ring-protection and ring-breaking channels are represented by blue and red colours,
     respectively. In the case of \pyrrolewi additional relaxation pathways emerged through the
     cleavage of the hydrogen bond for the first dissociation band. For simplicity, only one of the
     observed fragments is shown for each dissociation pathway.}
  \label{fig:protection}
\end{figure}

We estimated the ring-fragmentation probability individually for pyrrole and \pyrrolew after single
ionisation to 31~\% and 6~\%, respectively. \autoref{fig:protection} provides a sketch of the
corresponding potential-energy surfaces of \pyrrolei and \pyrrolewi with the resulting products and
their probabilities.
Single ionisation of pyrrole into the ground and first excited state of the cation led to the
formation of a stable parent ion, \pyrrolei~\cite{Rennie:CP250:217, VanDenBrom:PCCP7:892}. The
ionised molecule underwent fragmentation through the higher excited cationic
states~\cite{Palmer:CP238:179}, whose probability was 31~\%. In the case of \pyrrolew, single
ionisation into the ground state resulted in a stable parent ion,
\pyrrolewi~\cite{Schuetz:PCCP19:3970}. The first excited state of \pyrrolewi dissociated the
heterodimer into a \pyrrolei and neutral water. For the higher excited states of \pyrrolewi a
reduction in the ring fragmentation probability was observed. This was due to dissociative channels,
where the water is leaving neutrally. Obviously, for the pyrrole monomer such channels do not exist
after single ionisation.

Overall, the ring-fragmentation probability of pyrrole following single ionisation is reduced by a
factor of approximately $31/6=5.2$ in \pyrrolew compared to pyrrole. Based on the observation of
reduced fragmentation for \pyrrolewi and the assumption of a similar cross section
for single ionisation, we can infer that neutral molecules can be protected against radiation damage
through outer-valence ionisation by solvation even with a specifically-bound single water molecule.

To shed light on radiation damage effects after double ionisation, we classified the channels
following double ionisation of pyrrole and \pyrrolew. For pyrrole, the only ring-protecting channel
was the formation of a (meta)stable doubly-charged parent ion, \pyrdii, marked by the asterisks in
the upper panel of \autoref{fig:compare}. All other channels broke the aromatic ring. The dominant
ring protection channel for the \pyrrolewdii was the intermolecular proton transfer from the N--H
site of the pyrrole moiety to the water moiety, producing \hydron and \pyrmhi. A second channel was
an electron transfer process across the hydrogen bond, which led to the formation of \pyri and
\HHOi. All other double-ionisation channels broke the aromatic ring, but in our experiment these
were minor contributions. The ring-fragmentation probability for pyrrole and \pyrrolew after double
ionisation was determined by counting ions in the high-KE part of the momentum-map images while
taking into account that two ions might have been produced after double ionisation, leading to two
hits in the corresponding momentum maps for a single fragmentation event; the finite detection
efficiency of $0.5$ for each ion was statistically taken into account. This led to a similar
ring-fragmentation probability after double ionisation for \pyrrolew, $P=0.72\pm0.04$ and pyrrole,
$P=0.78\pm0.04$. Furthermore, we extracted a similar double-ionisation cross-section within our
model, see \methods. This observation is consistent with our previous assumption that the cross
sections for single ionisation are similar for both species.

\section{Conclusion}
We demonstrated that a single water molecule strongly protects the pyrrole molecule from
fragmentation after single ionisation. Furthermore, we observed similar ring-fragmentation
probabilities for \pyrrole and \pyrrolew in the case of double ionisation. These quantitative
studies were enabled by the production of pure samples of the bare molecule as well as the
singly-microsolvated complex and by the use of the versatile Timepix3 detector.

Singly-ionised pyrrole underwent radiation-induced damage through the breaking of, typically two,
C--C or N--C bonds. Notably, the dissociation mechanisms after single ionisation of \pyrrolew,
through breaking of the intermolecular bond or by transferring an electron or proton across this
hydrogen bond, strongly reduces the ring breaking probability by a factor of $5.2$. We inferred that
solvation effects also provided radiation protection for the neutral \pyrrolew system.

The radiation protection mechanisms for excited chromophores were studied both experimentally and
theoretically~\cite{Hertel:RPP69:1897, Domke:NatChem5:1755, Sobolewski:PCCP4:1093,
   Sobolewski:CP259:181, Onvlee:NatComm13:7462}. Our study showed that this protective effect
observed for solvated chromophores is also valid after ionisation.

Following double ionisation similar ring-fragmentation probabilities were observed for pyrrole and
\pyrrolew. In the microsolvated system, intermolecular proton- and electron transfer processes
occurring across the hydrogen bond increased the redistribution of charges, initially created in the
pyrrole ring, to the water molecule. This charge-distribution was observed through the formation of
\HHOi and \hydron, where the counter ion is either an intact ring or a ring-fragmentation product.

We employed strong-field ionisation for the removal of electrons from the low-binding-energy
molecular orbitals, which are localised on the pyrrole moiety in the monomer as well as the
pyrrole-water heterodimer. This process mimics the ionisation by slow secondary
electrons in aqueous systems, such as cells, as well as ionisation through VUV radiation -- which
both typically create singly-charged molecules since these ionisation prosesses access the same
potential energy surfaces and therefore dissociative states.

Our results provide a test case of how an aqueous microsolvation environment can strongly reduce the
radiation damage of biological molecules induced by ionising radiation as well as the biologically
important process of secondary effects of ionising radiation, where single outer-valence ionisation
of the biomolecular chromophore is the scenario. Furthermore, the doubly-ionised systems in our
experiment resemble to a large extent the fate of a molecule after Auger decay processes subsequent
to direct core-shell ionisation~\cite{Jahnke:JPB48:082001, Cederbaum:PRL79:4778}.

Biomolecules and proteins in nature are actively solvated by the surrounding water molecules, which
allow for efficient charge redistribution to the solvent environment through electron- and proton
transfer pathways quantified here. In aqueous solution, the loss of the attached -- neutral,
ionised, or protonated -- water could easily be repaired by the many solvent molecules around. Our
analysis of the protection in \pyrrolew provides a quantitative analysis of radiation protection and
serves as the basis for further detailed investigations regarding the role of the solvent
environment on the radiation damage of biomolecules.

\section{Methods}
\label{sec:methods}
\subsection{Experimental Setup}
Details of the experimental setup were described elsewhere~\cite{Trippel:MP111:1738}. A pulsed valve
was operated at 100~Hz to supersonically expand a few millibars of pyrrole (Sigma Aldrich, $>98$~\%)
and traces of water in $\ordsim90$~bar of helium into vacuum. The resulting molecular beam contained
atomic helium, individual pyrrole and water molecules, and various aggregates thereof. The electric
deflector, which enables the spatial separation of neutral polar molecules according to their
dipole-moment-to-mass ratio~\cite{Filsinger:PRL100:133003, Chang:IRPC34:557}, was used to create
pure samples of \pyrrolew with a typical purity close to 100\%~\cite{Johny:CPL721:149}. Pyrrole and
\pyrrolew were strong-field ionised by 800~nm laser pulses with linear polarisation, a duration of
$\ordsim30$~fs, focused to $\varnothing\approx35~\um$ (full width at half maximum intensity) with a
peak intensity of $\ordsim1\times10^{14}$~\Wpcmcm. The ions generated were extracted perpendicular
to the molecular beam and laser propagation directions using a velocity-map-imaging spectrometer
(VMIS). All ions were detected using a position- and time-sensitive detector consisting of a
micro-channel plate (MCP) in combination with a fast phosphor screen (P46). A
visible-light-sensitive Timepix3 detector~\cite{Zhao:RSI88:113104, Roberts:JInst14:P06001} in an
event-driven mode recorded all signals, which were stored and centroided using our homebuilt pymepix
software~\cite{AlRefaie:JINST14:P10003}.

\subsection{Normalisation of the mass spectra}
\label{sec:normalisation}
\begin{figure}
   \includegraphics[width=\linewidth]{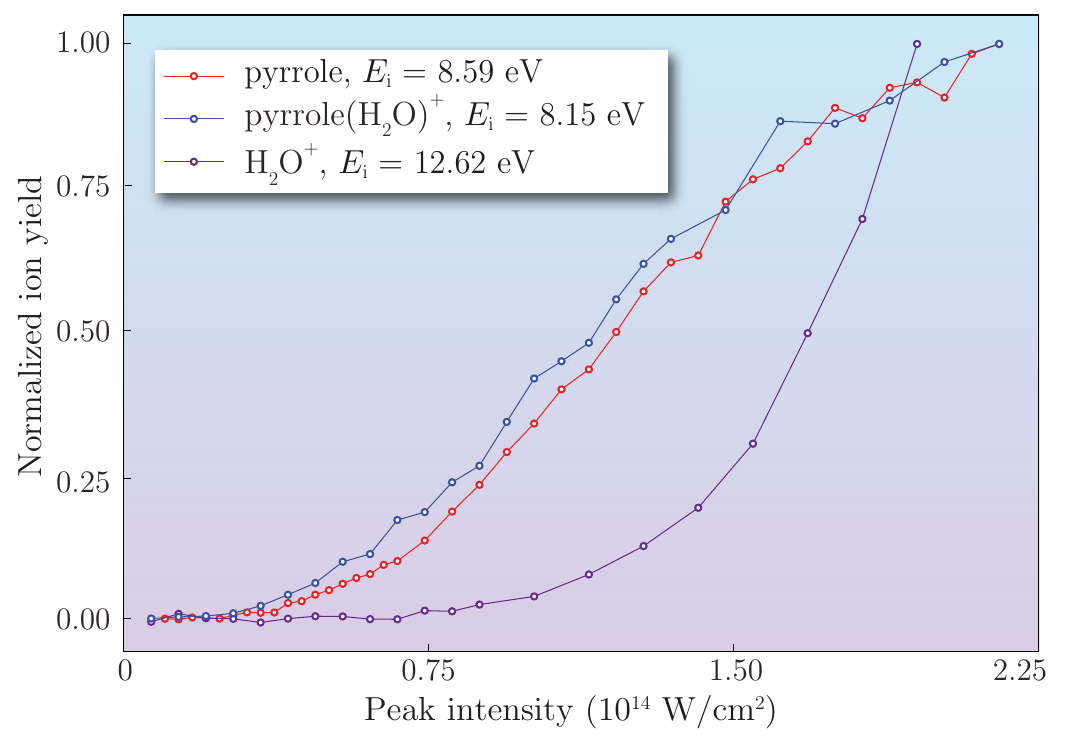}%
   \caption{Single-ionisation yields for pure samples of pyrrole (red), \pyrrolew (blue), and water
      (purple) measured at different laser peak intensities. All curves are normalised to their
      highest ion yield observed.}
   \label{fig:tof_normalisation}
\end{figure}
The normalisation of the TOF-MS was necessary due to the different densities of the two species,
pyrrole and \pyrrolew, in the molecular beam. Due to the very similar first ionisation energies
(\Ei) of pyrrole and \pyrrolew the ionisation probabilities for both species are also very similar.
The calculated (HF/MP2-aug-cc-pVTZ using GAMESS-US) first vertical \Ei of pyrrole and \pyrrolew are
$8.59$~eV and $8.15$~eV, respectively. To quantify the relative ionisation probability
experimentally~\cite{Hankin:PRA64:013405, Wiese:NJP21:083011} the ion yield as a function of the
laser peak intensity was measured for the single-ionisation channel for pyrrole, \pyrrolew, and
water, see \autoref{fig:tof_normalisation}. The spectrometer was defocussed in terms of VMI
conditions to avoid saturation of the detector. The ion yield from each pure species is normalised
to its value for the highest laser intensity, demonstrating that the peak-intensity-dependent shape
of the curves is indeed very similar, confirming the similarity of the
\Ei~\cite{Hankin:PRA64:013405}. In the low-intensity region, $1\text{--}8\times10^{13}$~\Wpcmcm, for
pyrrole only the parent ion was observed. A saturation intensity for the parent ion signal of
$\ordsim6\times10^{13}$~\Wpcmcm was obtained from the measured low-intensity ion-yield curve. For
\pyrrolew we summed the signals for parent ion and \pyrrolei~\cite{Johny:CPL721:149}, which yielded
a saturation intensity for single ionisation of $\ordsim4.5\times10^{13}$~\Wpcmcm. Based on the
similar saturation intensities and the very comparable intensity dependence of the ionisation
yields, \autoref{fig:tof_normalisation}, we assumed similar single-ionisation cross-sections and
normalised the TOF-MS of pyrrole and \pyrrolew using a normalisation factor of $3.33\pm0.1$. This
normalisation factor is identical to the relative number densities of pyrrole and \pyrrolew beam.

For the ionisation of water, we obtained a very different intensity dependence and saturation
intensity, which was determined as $\ordsim1.4\times10^{14}$~\Wpcmcm. This is consistent with the
larger \Ei of $12.62$~eV~\cite{Page:JCP88:5362} and further demonstrates the similarities in the
ionisation cross sections of pyrrole and \pyrrolew.

\subsection{Highest occupied molecular orbitals of \pyrrolew}
\label{sec:molecular_orbitals}
\begin{figure}
   \includegraphics[width=1\linewidth]{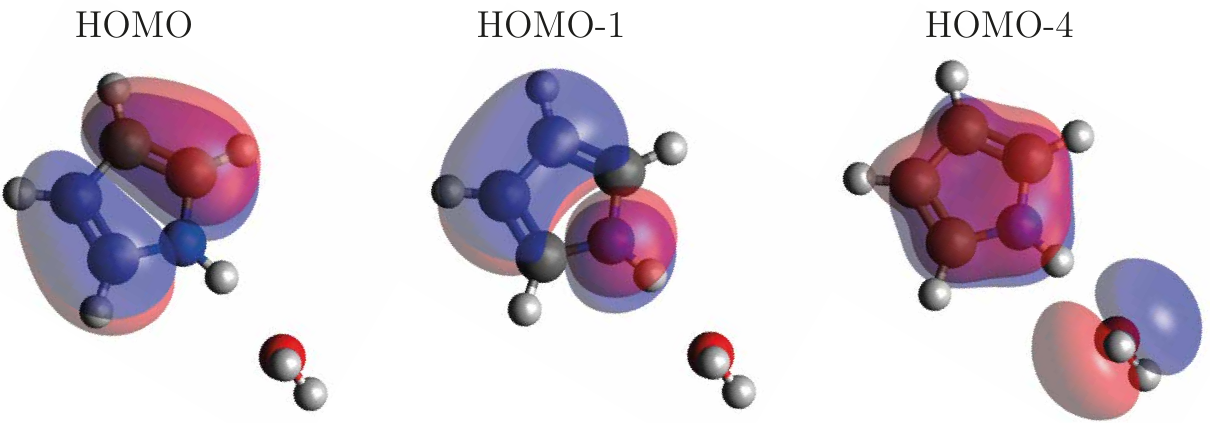}%
   \caption{Molecular orbital picture of HOMO, HOMO-1, and HOMO-4 for the geometry optimised ground
     state structure of the \pyrrolew.}
   \label{fig:orbital}
\end{figure}
The molecular orbitals calculated (GAMESS-US, HF/MP2-aug-cc-pVTZ) for the geometry-optimised ground
state structure of \pyrrolew are shown in \autoref{fig:orbital}. The electron densities of HOMO to
HOMO-3 are localised on the aromatic ring. The highest-energy bound molecular orbital with
significant density on the water moiety is HOMO-4. This orbital has an energy that is 6.5~eV lower
than the HOMO. Therefore, under the applied laser intensities ionisation from this orbital can be
neglected~\cite{Popruzhenko:JPB47:204001} and localised ionisation of \pyrrolew at the pyrrole
moiety can be safely assumed.

\subsection{Double-ionisation yield}
We compared the absolute total ion yield of the monomer and the singly-microsolvated system after
double ionisation. This double-ionisation ion yield was indicated qualitatively through the
comparison of the ion signals in the NORMS in \autoref{fig:compare}. However, a direct comparison of
the yields in the TOF-MS was not straightforward due to the complex fragmentation pathways and the
overlap of higher-order ionisation channels. Therefore, to quantify this, we counted the ions in the
normalised momentum maps of the specific channels originating from double ionisation of pyrrole and
\pyrrolew. The normalisation was done using the extracted relative number density ratios between
these two species, which was based on the assumption of similar single-ionisation cross-section for
both species. The direct comparison of the total double-ionisation yields revealed a reduction of
the total ion yield of \pyrrolew, with respect to pyrrole, by a factor of $1.22\pm0.3$. Thus, we
concluded that the double-ionisation cross-section is also similar for pyrrole and \pyrrolew within
the errors of our estimation model.

\subsection{Triple-ionisation contributions}
\begin{table*}
   \centering%
   \begin{tabular}{ccccc}
     \hline\hline
     radius $p_r$ & ion counts in disk & ion counts in ring & area of disk/$\pi$ & area of ring/$\pi$ \\
     \hline
     $60$ & $167522$ & $167522$ & $3600$ & $3600$ \\
     $140$ & $225860$ & $58338$ & $19600$ & $16000$ \\
     $200$ & $229120$ & $3260$ & $40000$ & $20400$ \\
     \hline\hline
   \end{tabular}
   \caption{Total ion counts as well as the areas of disks and rings within specific radii chosen in
      the momenta map.}
   \label{tab:ioncount1}
\end{table*}
\begin{table*}
   \centering
   \begin{tabular}{ccc}
     \hline\hline
     charge state & ion counts & ion counts/area \\
     \hline
     +1 (single) & $167522-3.486\cdot3600-0.1598\cdot3600 = 154397$ & $154397/3600=42.89$ \\
     +2 (double) & $58338\cdot(19600/16000) -0.1598\cdot19600 = 68332$ & $68332/19600=3.486$ \\
     +3 (triple) & $3260\cdot(40000/20400) = 6392$ & $6392/40000 = 0.1598$ \\
     \hline\hline
   \end{tabular}
   \caption{Number of ion counts per area of each disk, and the estimated number of ions formed
      after the single, double, and triple ionisation, respectively.}
   \label{tab:ioncount2}
\end{table*}
\begin{figure}[t]
   \centering%
   \includegraphics[width=1\linewidth]{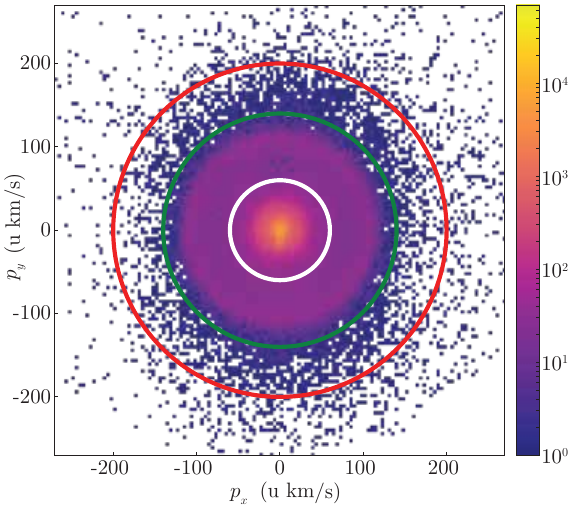}%
   \caption{The momentum map for all ions detected within a mass-to-charge region \mqr{35}{45} is
      shown. Marked circles with specific radii in the momenta map represent edges for single,
      double, and triple ionisation, respectively.}
   \label{fig:triple_estimation_momenta}
\end{figure}
To estimate the contributions of individual ions over a given mass-to-charge range from the measured
2D projection of the 3D momenta, we made specific cuts in these experimental momentum maps in order
to isolate contributions from single, double, and triple ionisation processes.

The measured momentum map for \mqr{35}{45} is shown with \autoref{fig:triple_estimation_momenta} as
an example. Circles indicate corresponding cuts in the 2D projection of the 3D momentum sphere
formed from each ionisation process: The white circle with a radius of $p_r=60~\ukms$ represents the
edge of the momentum for dissociative single ionisation, the green circle with $p_r=140~\ukms$
corresponds to the maximum momentum of ions from Coulomb explosion following the double ionisation,
and the red circle with $p_r=\sqrt{2}*140\approx200~\ukms$ is the maximum momentum of ions from
triple ionisation, assuming a two-body fragmentation into a singly-charged and a doubly-charged ion.

The total ion count corresponding to the disks defined by these specific radii and the signal in the
two outer rings are provided in \autoref{tab:ioncount1}. Areas of the specific disks and rings are
also given. Ion counts inside the outer ring, $140<p_r<200$, correspond to triple ionisation without
contribution from single and double ionisation. However, the middle ring, $60<p_r<140$, represents
the double-ionisation channels and it has contributions from triple ionisation; similarly, the
innermost disk, $0<p_r<60$, representing single ionisation has contributions from ions originating
in double and triple ionisation. The corrected total number of ions from single, double, and triple
ionisation are provided in \autoref{tab:ioncount2} assuming a flat ion distribution in the inner
part of the corresponding rings and disks. The relative contribution of the ion yield from the
triple-ionisation process to the total ion yield in the given mass-to-charge region is $<5$~\%, \ie
negligible.

\subsection{Determination of real ion numbers taking into account detector saturation}
Here, the connection between the measured number of ions \Ndet and the real number of ions produced
\Nreal taking into account saturation effects of the detection system are discussed. The saturation
effects appear, \eg for the parent ions, pyrrole and pyrrole-water, due to the small areas on the
detector where these ions are detected. The small areas are accounted for the fact that the
spectrometer was operated in VMI conditions in combination with the translational molecular beam
temperature below 1~K. Furthermore, all parent ions arrive on the detector at the same TOF within
the temporal resolution of the Timepix3 camera. This makes it impossible to determine the real
number of parent ions per shot by counting since only a single event is detected at an instant of
time by our detection system. Assuming a Poissonian distribution $P_\lambda(\Nreal)$ of the real
ions created, however, allows to estimate the real number of ions \Nreal from the detected number of
ions \Ndet. \autoref{tab:single} summarises the numbers and probabilities of various channels.
\begin{table*}
  \begin{center}
    \begin{tabular}{|p{2.5cm}|p{2.5cm}|c|c|c|c|c|c|}
      \hline
      Parent molecule & Ion        & \Ndet  & \Nshots & \Rdet   & $P_\lambda(0)$ & $\lambda$ & \Nreal  \\
      \hline
      \pyrrole        & \pyrrolei  & 234626 & 270028  & 0.8688  & 0.1312        & 2.0310    & 548485  \\
      \hline
      \pyrrole        & fragments   & 241298 & 270028  & 0.89360 & -             &  -        & 241298  \\
      \hline
      \pyrrolew       & \pyrrolei  & 926658 & 1801197 & 0.5144  & 0.4856        & 0.7223    & 1301176 \\
      \hline
      \pyrrolew       & \pyrrolewi & 152728 & 1801197 & 0.08479 & 0.91521       & 0.08860   & 159591 \\
      \hline
      \pyrrolew       & fragments   & 91799  & 1801197 & 0.05097 & -             & -         & 91799  \\
      \hline
    \end{tabular}
    \caption{\label{tab:single} Measured number of ions, probabilities, correction factors, and real
       number of ions.}
  \end{center}
\end{table*}
The first column indicates the parent system under investigation. The corresponding ions are listed
in the second column including the fragments. The corresponding number of laser shots are denoted
with \Nshots. The detected hit rate per shot \Rdet = \Ndet/\Nshots indicates the probability to
detect an event for a single laser shot per channel. The probability to detect 0 ions within a laser
shot is given by $P_\lambda(0)=1-R_\mathrm{det}$. This value can be used to determine the mean
number of ions per shot $\lambda = -\ln(P_\lambda(0))$ according to the Poissonian distribution. The
real number of ions \Nreal is then given by $\Nreal=\Nshots\lambda$. For the ring fragmentation
channels, we set \Nreal = \Ndet due to the larger spatio-temporal volume covered by these ions on
the detector. The ring fragmentation probability for \pyrrole and \pyrrolew after SFI taking into
account saturation effects of the detection system is therefore given by
$P_\mathrm{P} = (548485+241298)/241298 = 3.273$ and
$P_\mathrm{PW} = (1301176+159591+91799)/91799 = 16.912$, respectively. This gives rise to a ring
protection factor given by $P = P_\mathrm{PW}/P_\mathrm{P} = 16.912/3.273 = 5.167$.

\subsection{Pyrrole and pyrrole-water isotopologue peaks}
The missing isotopologue peak for pyrrole is attributed to the dead time in the order of \us of the
Timepix3 camera in combination with the relatively high detection rate per shot given by
$\Rdet=0.8688$ (see \autoref{tab:single}) and the finite spatio-temporal resolution. The natural
abundancy of C$_{13}$ isotope is in the order of 1.1\%. This results in an expected relative peak
height in the order of 4.5\% for the C$_{13}$ parent ion peak due to the 4 carbon atoms present in
pyrrole. Due to the dead time, we expect only C$_{13}$ containing parent ions to be detected when no
C$_{12}$-only containing parent ion was detected before. Taken into account the ion rate per shot
provided for pyrrole, results in a dead time corrected relative peak height for its isotopologue of
$(1-0.8688)\cdot0.045 = 0.6\%$. This is in the noise level of the pyrrole peak. For the doubly
charged parent, ion the isotopologue peak is present in ~\autoref{fig:pyr:tof}. Here, we only have a
detection rate per shot given by 0.1073. The rate for the corresponding isotopologue peak is given
by 0.00381. This gives rise to a corresponding C$_{13}$ containing fraction of 3.4\% close to the
estimate provided above. The same arguments hold for the observed isotopologue peaks in the case of
pyrrole-water ~\autoref{fig:pyrwat:tof}. Here, due to lower count rates, the isotopologue peaks are
present for \pyri and \pyrwi.

\subsection*{Data and materials availability}
The data that support the ﬁndings of this study are available from the repository at
\href{https://doi.org/10.5281/zenodo.7857990}{https://doi.org/10.5281/zenodo.7857990}

\subsection*{Code availability}
The code used for the data analysis is available at
\href{https://doi.org/10.5281/zenodo.7857990}{https://doi.org/10.5281/zenodo.7857990}

\bibliographystyle{unsrtnat}
\bibliography{string,cmi}

\subsection*{Acknowledgments}
We thank Jolijn Onvlee, Jonathan Correa, and David Pennicard for helpful discussions.

\subsection*{Funding}
We acknowledge financial support by Deutsches Elektronen-Synchtrotron DESY, a member of the
Helmholtz Association (HGF), the European Union's Horizon 2020 research and innovation program under
the Marie Skłodowska-Curie Grant Agreement 641789 ``Molecular Electron Dynamics investigated by
Intense Fields and Attosecond Pulses'' (MEDEA), the Clusters of Excellence ``Center for Ultrafast
Imaging'' (CUI, EXC~1074, ID~194651731) and ``Advanced Imaging of Matter'' (AIM, EXC~2056,
ID~390715994) of the Deutsche Forschungsgemeinschaft (DFG), and the European Research Council under
the European Union's Seventh Framework Programme (FP7/2007-2013) through the Consolidator Grant
COMOTION (614507). We acknowledge the use of the Maxwell computational resources operated at
Deutsches Elektronen-Synchrotron DESY, Hamburg, Germany.

\subsection*{Author contributions}
M.J., S.T., and J.K.\ conceived the experiment; M.J., C.S., A.R., L.H., and S.T.\ performed the
experiment; A.R.\ adopted the pymepix software; M.J.\ and S.T.\ analysed the data; M.J., J.W., S.T.,
and J.K.\ interpreted the results; M.J., S.T., and J.K.\ prepared the manuscript. All authors
discussed the results and the manuscript.

\subsection*{Competing interests}
The authors declare that they have no competing interests.

\onecolumngrid
\listofnotes
\end{document}